\documentclass[onecolumn,preprint]{revtex4}
\usepackage{graphicx}
\usepackage{times}

\begin{document}

\title{Electrically-driven phase transition in magnetite nanostructures}

\author{Sungbae Lee$^{1}$, Alexandra Fursina$^{2}$, John T. Mayo$^{2}$, Cafer T. Yavuz$^{2}$,
Vicki L. Colvin$^{2}$, R.~G. Sumesh Sofin$^{3}$, Igor V. Shvets$^{3}$, Douglas Natelson$^{1, 4}$}

\affiliation{$^{1}$ Department of Physics and Astronomy, Rice University, 6100 Main St., Houston, TX 77005, $^{2}$ Department of Chemistry, Rice University, 6100 Main St., Houston, TX 77005, $^{3}$ CRANN, School of Physics, Trinity College, Dublin 2, Ireland, $^{4}$ Department of Electrical and Computer Engineering, 
Rice University, 6100 Main St., Houston, TX 77005}


\begin{abstract} 
{\bf Magnetite (Fe$_3$O$_4$), an archetypal transition metal oxide, has
been used for thousands of years, from lodestones in primitive
compasses\cite{Mills:2004} to a candidate material for
magnetoelectronic devices.\cite{Coey:2003} In 1939
Verwey\cite{Verwey:1939} found that bulk magnetite undergoes a
transition at $T_{\mathrm{V}} \approx 120$~K from a high temperature
``bad metal'' conducting phase to a low-temperature insulating phase.
He suggested\cite{Verwey:1941} that high temperature conduction is via
the fluctuating and correlated valences of the octahedral iron atoms,
and that the transition is the onset of charge ordering upon cooling.
The Verwey transition mechanism and the question of charge ordering
remain highly
controversial.\cite{Walz:2002,Garcia:2004,Huang:2006,Nazarenko:2007,Subias:2004,Rozenberg:2006,Piekarz:2006}
Here we show that magnetite nanocrystals and single-crystal thin films
exhibit an electrically driven phase transition below the Verwey
temperature.  The signature of this transition is the onset of sharp
conductance switching in high electric fields, hysteretic in voltage.
We demonstrate that this transition is not due to local heating, but
instead is due to the breakdown of the correlated insulating state
when driven out of equilibrium by electrical bias.  We anticipate that
further studies of this newly observed transition and its
low-temperature conducting phase will shed light on how charge
ordering and vibrational degrees of freedom determine the ground state
of this important compound.}
\end{abstract}

\maketitle



Strongly correlated electronic materials can exhibit dramatic
electronic properties ({\it e.g.}, high temperature superconductivity,
metal-insulator transitions, and charge ordering) not present in
simple systems with weaker electron-electron interactions.  Such rich
electronic phenomenology can result when electron-electron
interactions, electron-phonon interactions, and electronic bandwidth
are all of similar magnitude, as in magnetite.\cite{Gasparov:2000}
Verwey\cite{Verwey:1939} found nearly seven decades ago that bulk
magnetite, while moderately conductive at room temperature, undergoes
a transition to a more insulating state below what is now called the
Verwey temperature, $T_{\mathrm V} \approx 120$~K.  Similar
transitions are known in a number of
materials.\cite{Imada:1998,Coey:2004} Above $T_{\mathrm V}$,
Fe$_{3}$O$_{4}$ has an inverse-spinel structure of the form
AB$_{2}$O$_{4}$, with tetrahedrally coordinated A sites occupied by
Fe$^{3+}$ and octrahedrally coordinated B sites of mixed valence,
equally occupied by irons with formal $+3$ and $+2$ charges.
Conduction at high temperatures has long been thought to be through
fluctuating valences of the B sites, with the transition corresponding
to some kind of B site charge ordering as $T$ decreases; concurrent is
a first-order structural phase transition to an orthorhombic unit
cell.  This explanation remains
controversial,\cite{Walz:2002,Garcia:2004} with experiments showing
some charge disproportion or charge order
(CO),\cite{Huang:2006,Nazarenko:2007} and others implying that the
structural degrees of freedom drive the change in
conductivity.\cite{Subias:2004,Rozenberg:2006} Recent theoretical
progress has been made in understanding the complex interplay of
charge and structural degrees of freedom\cite{Leonov:2006,Pinto:2006},
including a complete picture of the transition mechanism
\cite{Piekarz:2006} with strongly correlated 3$d$ Fe electrons acting
to amplify electron-phonon couplings.  Testing these ideas
experimentally is of much interest.

In this Letter we report electronic transport measurements in
magnetite at the nanoscale on both nanocrystals and single-crystal
epitaxial thin films.  Both types of devices exhibit striking
electrically-driven hysteretic switching of the electronic conductance
once sample temperatures are reduced below $T_{\mathrm V}$.  The data
clearly show that the transition is not the result of local heating
above $T_{\mathrm V}$, but instead is an electrically-driven breakdown
of the insulating state.  We discuss possible explanations for this
switching in the context of the general Verwey transition
problem.  While qualitatively similar resistive switching has been
observed in other correlated oxide
systems\cite{Asamitsu:1997,Sawa:2004}, the
phenomenon in Fe$_{3}$O$_{4}$ is a bulk effect with a mechanism
distinct from these.

\begin{figure}
\begin{center}
\includegraphics[clip, width=8cm]{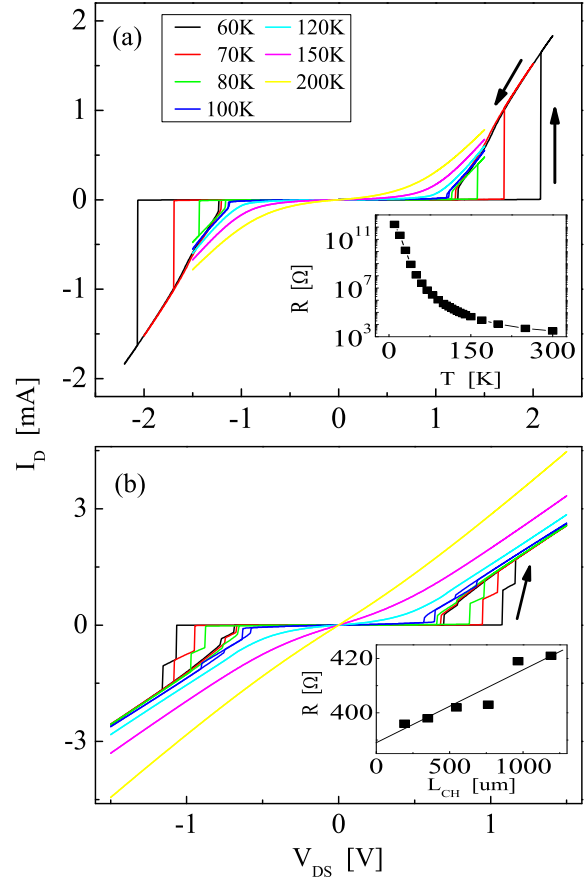}
\end{center}
\vspace{-5mm}
\caption{\small Hysteretic conductance switching below $T_{\mathrm
V}$.  (a) Current-voltage characteristics at various temperatures for
a device based on 10~nm magnetite nanocrystals.  Arrows indicate the
direction of the hysteresis.  Inset: Zero-bias resistance, $R(T)$. (b)
Analogous data for a device based on a 50~nm-thick MBE-grown magnetite
film.  The nominal interelectrode gap was planned to be 100~nm, but at
its narrowest was approximately 10~nm.  Inset: Two-terminal resistance
as function of channel length for another set of devices fabricated on
another piece of the same film.}
\label{fig1}
\vspace{5mm}
\end{figure}

Two-terminal devices for applying voltages and measuring conduction
at the nanoscale have been fabricated (see Methods) incorporating
both Fe$_{3}$O$_{4}$ nanocrystals\cite{Yu:2004} (10-20~nm in diameter
with oleic acid coating) and single-crystal thin films (40-60~nm
thick).\cite{Zhou:2004} Devices were measured in both a variable
temperature vacuum probe station and a $^4$He cryostat with magnet.
Current-voltage characteristics have been measured with both a
semiconductor parameter analyzer and directly using voltage sources
and current amplifiers, with differential conductance computed
numerically.


Figure 1a shows $I-V$ characteristics of a nanocrystal device at
selected temperatures.  When cooling, zero-bias conductance decreases
monotonically until $T\rightarrow T_{\mathrm V}$.  Below $T_{\mathrm
V}$, the $I-V$ characteristics show sharp switching between a low bias
insulating state and a high bias state with much higher differential
conductance $dI/dV(V)$ (close to $dI/dV(V=0, T=300~{\mathrm K}))$,
with dramatic hysteresis as a function of voltage sweep direction.
The switching threshold voltages increase in magnitude as $T$ is
decreased.

Dozens of nanocrystal devices were measured and only those with 300~K
resistances below 10~k$\Omega$ showed the switching, with higher
resistance devices having higher switching threshold voltages.
Resistances decrease by some three orders of magnitude with vacuum
annealing at 673~K, likely because of oleic acid
decomposition.\cite{Perezdieste:2003,Zeng:2006} The temperature
dependence (Fig. 1a, inset) of the zero-bias resistance, $R(T)$, has no step at
$T_{\mathrm V}$, showing that $R(T)$ remains dominated by contact
effects.  


\begin{figure}
\begin{center}
\includegraphics[clip, width=8cm]{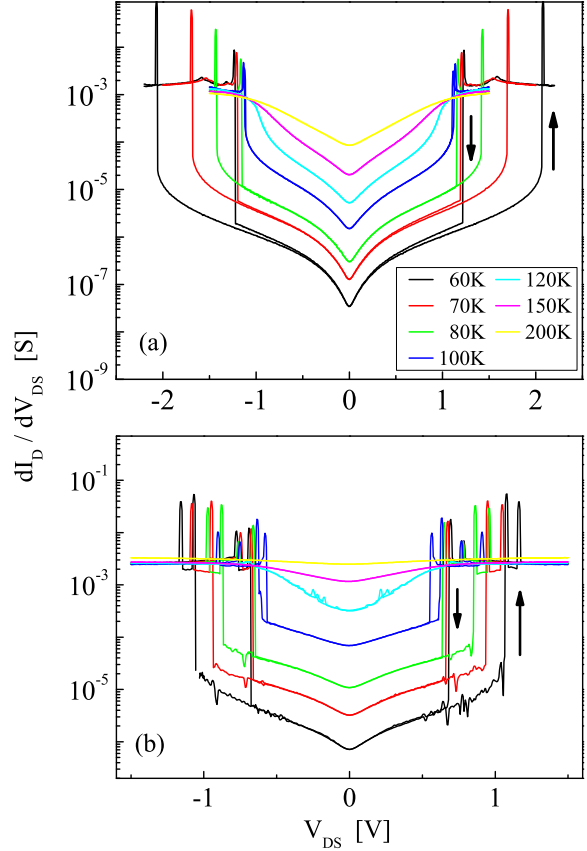}
\end{center}
\vspace{-5mm}
\caption{\small Differential conductance plots of the switching.  (a)
$dI/dV$ vs. $V$ for the nanocrystal device from Fig. 1a at the same
temperatures.  (b) $dI/dV$ vs. $V$ for the thin film device from
Fig. 1b.  Both plots have logarithmic $dI/dV$ axes to better show the
lowest temperature data. }
\label{fig2}
\vspace{5mm}
\end{figure}


Qualitatively identical conduction is apparent in the thin film
devices, as shown in Fig. 1b.  Contact resistances are also important
in these structures, as demonstrated by examining $R$ at low bias ($<
100$~mV) as a function of channel length, $L$, as shown in the inset
for one set of devices.  Extrapolating back to $L \rightarrow 0$, the
contact resistance, $R_{\mathrm c}$, at 300~K is 390~$\Omega$, while
the 50~nm thick channel of width 20~$\mu$m contributes
27.2~$\Omega$/micron, implying (based on channel geometry) a magnetite
resistivity of 2.9~m$\Omega$-cm, somewhat below bulk expectations.
Further investigations are seeking to understand and minimize
$R_{\mathrm c}$.  Analysis of $R(L)$ at lower temperatures shows that
$R_{\mathrm c}$ increases significantly as $T$ is decreased, exceeding
80~k$\Omega$ by 80~K.  This complicates the analysis of the switching,
since some of the total $V$ is dropped across $R_{\mathrm c}$ rather
than directly within the Fe$_{3}$O$_{4}$; further, the contacts may
not be Ohmic near the switching threshold.  We return to this issue
later.

The transitions in all devices are extremely sharp, with widths less
than 50~$\mu$V, though in repeated sweeps at a fixed temperature,
there is sweep-to-sweep variability of a few mV in switching
thresholds.  Using the substrate as a gate electrode, no discernable
gate modulation was seen in nanocrystal devices for gate biases
between {-80~V} and {+80~V}; this suggests that nanocrystal charging
effects do not dominate.  Switching characteristics were independent
of magnetic field perpendicular to the sample surface up to 9~T,
showing no large coupling between magnetization and the transition.

Differential conductance traces (Figure 2) show the transition even
more dramatically.  In the high conductance state, $dI/dV$ is
relatively temperature independent.  As $T$ is decreased, a clear
zero-bias suppression develops, deepening into a hard gap when $T <
T_{\mathrm V}$.  In the nanocrystal data there are indications (in
$d^{2}I/dV^{2}(V)$) of gap formation even at 150~K.  We note that
$T_{\mathrm V}$ in nanocrystals could be elevated, since nanocrystals
have large surface-to-volume ratios and the transition temperature of
the magnetite surface is known to be higher than in the
bulk.\cite{Shvets:2004}

\begin{figure}  
\begin{center} 
\includegraphics[clip, width=8cm]{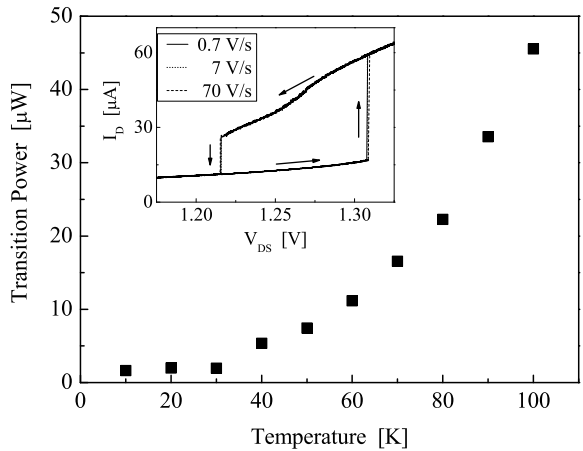} 
\end{center} 
\vspace{-5mm} 
\caption{\small Power required to switch from the insulating into the more
conducting state as a function of temperature, for a device based on $\sim$20~nm diameter nanocrystals.  Inset:  Hysteresis loop in the conduction of the same
device at 80~K, showing essentially no change in switching
characteristics as the voltage sweep rate is varied over two orders 
of magnitude. } 
 \label{fig3}
\vspace{5mm} 
\end{figure}


Several lines of evidence indicate that these sharp conductance
transitions are {\it not} the result of local Joule heating (as in
macroscopic samples of Fe$_3$O$_4$ \cite{Burch:1969,Freud:1969} and in
the Mott insulator VO$_{2}$ \cite{Duchene:1971,Gu:2007}), but rather
are electrically driven.  In the worst-case scenario, all of the $I
\times V$ Joule heating power is dissipated within the magnetite, and 
inhomogeneous dissipation ({\it e.g.}, filamentary conduction through
a locally heated path) can complicate the analysis.  The
local steady-state temperature is determined by the power dissipated
and the thermal path.  Thermally driven switching would then
correspond to raising the \textit{local} temperature above $T_{\mathrm
  V}$.  At a fixed cryostat temperature an improved thermal path would
imply that more power dissipation would be required for a given local
temperature rise.  Similarly, for a fixed thermal path, the necessary
dissipated power for thermal switching would approach zero as $T
\rightarrow T_{\mathrm V}$.  Furthermore, at a given cryostat
temperature thermally-driven switching would imply that the power
dissipated at the low-to-high conductance transition (needed to
raise the local temperature to $T_{\mathrm V}$) should be close
to that at the high-to-low conductance transition.

The thermal conductivity, $\kappa$, of magnetite is dominated by
phonons in this temperature range, and limited by phonon-electron
scattering,\cite{Salazar:2004} even when $T>T_{\mathrm V}$.  As a
result, $\kappa$ increases as $T$ is \textit{decreased} through and
below $T_{\mathrm V}$, and the material's thermal coupling to the
cryostat \textit{improves} as $T$ is reduced.  In {\it all} devices
showing switching, the electrical power required to switch from low to
high conductance {\it decreases} with decreasing $T$, with Fig. 3
showing one example.  This is precisely the opposite of what one would
expect from thermally-driven switching.  Similarly, in all devices the
power dissipated at switching does {\it not} approach zero as $T
\rightarrow T_{\mathrm V}$, again inconsistent with thermally-driven
switching.  Furthermore, at a given $T$ the power dissipated just
before $V$ is swept back down through the high-to-low conductance
threshold significantly exceeds that dissipated at the low-to-high
point in many devices, including those in Fig.~1, inconsistent with
thermal switching expectations.  Finally, nanocrystal and thin film
devices show quantitatively similar switching properties and trends
with temperature, despite what would be expected to be very different
thermal paths.  These switching characteristics are also qualitatively
very different from those in known inhomogeneous Joule
heating.\cite{Burch:1969,Freud:1969}.  These facts rule out local
heating through the Verwey transition as the cause of the conductance
switching.

Figure 3 (inset) shows details of hysteresis loops on a nanocrystal
device comparing different voltage sweep rates.  The loop shape and
switching voltages are unchanged to within the precision of the data
collection as voltage sweep rates are varied from around 0.7~V/s up to
70~V/s.  This indicates that the switching process is relatively
rapid.  Further studies will examine the intrinsic switching speed.

The observed conductance transition appears to be driven {\it
  electrically}.  Figure 4 is a plot of the low-to-high conductance
switching voltage as a function of $L$ in a series of film devices for
several temperatures.  The linear dependence implies that the
transition is driven by electric field itself, rather than by the
absolute magnitude of the voltage or the current density.  
The fact that the voltage extrapolates to a nonzero value
at $L=0$ is likely a contact resistance effect.  Minimizing and better
understanding the contact resistance will allow the determination of
the electric field distribution within the channel.

The length scaling of the transition voltage also demonstrates that
this is a \textit{bulk} effect.  The contacts in all of these devices
are identical, so any change in switching properties must result from
the magnetite channel.  This is in contrast to the resistive switching
in Pr$_{0.7}$Ca$_{0.3}$MnO$_{3}$ (PCMO) that is ascribed to a change
in contact resistance due to occupation of interfacial
states\cite{Sawa:2004}.

\begin{figure}  
\begin{center} 
\includegraphics[clip, width=8cm]{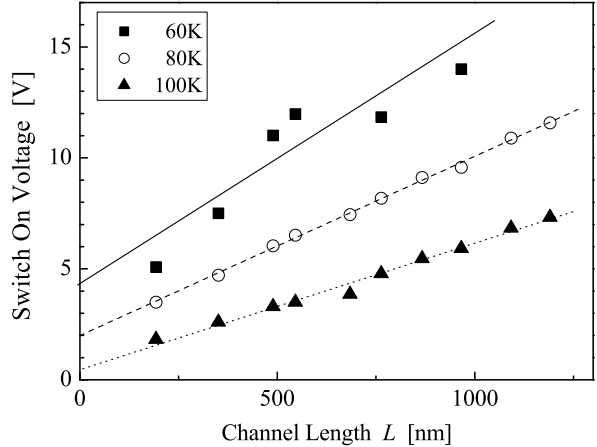} 
\end{center} 
\vspace{-5mm} 
\caption{\small Switching voltages in a series of film devices as a
function of channel length at several temperatures.  The linear
variation with $L$ strongly implies that for each temperature there is
a characteristic electric field required for switching.  The non-zero
intercepts of the trend lines indicate that some device-dependent
threshold voltage must be exceeded for switching even when $L=0$,
suggestive of contact effects.}
\label{fig4} 
\vspace{5mm}
\end{figure}


The field-driven conductance transition may give insights into the
equilibrium Verwey transition.  This switching may be useful in
testing recent calculations\cite{Piekarz:2006,Leonov:2006,Pinto:2006}
about the role of strongly correlated B-site Fe 3$d$ electrons and
their coupling to phonons in the Verwey transition mechanism.  It is
interesting to ask, to what degree is the field-driven electronic
transition coupled to the local structure?  It is greatly desirable to
perform local probes of the magnetite structure (via x-ray or
electron-diffraction techniques or scanned probe microscopy)
\textit{in situ} in the channel of biased devices, to see if the
coherence between structural symmetry changes and the formation of a
gap in the electronic spectrum is broken under these noequilibrium
conditions.  This is a significant experimental challenge.  Similarly,
local Raman spectroscopy of devices under bias could reveal
field-induced changes in phonon modes and electron-phonon couplings,
and single-crystal thin films permit the application of bias along
well-defined crystallographic directions relevant to structural
symmetry changes at $T_{\mathrm V}$.  We do note that qualitatively
identical switching occurs in nanocrystal devices as in strained thin
films strongly coupled to rigid MgO substrates.  This suggests that
elastic constraints on scales much larger than the unit cell have
relatively little influence on the observed switching.


It is also possible that the nonequilibrium carrier distribution
contributes to destabilizing the insulating state.  In the presence of
a strong electric field a carrier can gain significant energy even in
a {\it single} hopping step, even though carrier relaxation times are
very short.  A rough estimate of the average critical $E$-field for
switching at 80~K is $10^7$~V/m, from the slope of the line in Fig. 4.
The high temperature cubic unit cell is 0.84~nm on a side, meaning
that a carrier traversing one cell would gain approximately 8.4~meV,
comparable to $k_{\mathrm B}T_{\mathrm V} \approx$~10.3~meV.
Conductance switching at such high fields may require consideration of
such nonequilibrium carrier dynamics.

The presence of multiple switching transitions in individual
nanocrystal and film devices also bears further study.  The suggested
charge order may melt inhomogeneously, with portions of the channel
having different switching thresholds.  There could also be
charge-ordered intermediate states between the insulating regime and
the most conducting regime.\cite{Walz:2002} Again, optical
measurements\cite{Gasparov:2000} with sufficient spatial resolution
could address these possibilities.  Through improved metal/magnetite
contacts and further study, it should be possible to unravel the
precise nature of this nonequilibrium transition, and its relationship
to the equilibrium, bulk Verwey transition.


\section{Methods}

Magnetite nanocrystals were prepared via solution-phase decomposition
of iron carboxylate salts.\cite{Yu:2004}  The nanocrystals have been
characterized by transmission electron microscopy (TEM), x-ray 
diffraction, and infrared and Raman spectroscopy, as discussed
in Supplemental Material.  As synthesized the nanocrystals are 
protected by weakly bound oleic acid ligands; these ligands
allow the suspension of the nanocrystals in organic solvents, 
but act as electrically insulating layers that must be largely
removed for effective electronic transport measurements.  

Two-step electron beam lithography and e-beam evaporation (1~nm Ti,
15~nm Au) were used to pattern closely spaced source and drain
electrode pairs onto degenerately $n$-doped silicon substrates coated
with 200~nm of thermally grown SiO$_{2}$.  Interelectrode gaps
(channel lengths) ranged from zero to tens of nm, with a 10~$\mu$m
wide channel region.  Nanocrystals were spin-coated from hexane
solutions to form slightly more than one densely packed monolayer of
nanocrystals over the channel region.  Samples were then baked
at 673~K in vacuum for 1~hr to remove as much of the 
oleic acid as possible.  In one set of samples, a second layer
of particles was added followed by a second round of baking.

The other class of devices are based on epitaxial magnetite films
50~nm thick grown by oxygen-plasma-assisted molecular beam epitaxy
(MBE) on $\langle 100 \rangle$ MgO single-crystal substrates.  Details
of the growth process have been reported elsewhere.\cite{Zhou:2004}
Single-step e-beam lithography and e-beam evaporation were used to
pattern Au (no Ti adhesion layer) source and drain electrodes defining
a channel length ranging from tens of nm to hundreds of nm, and a
channel width of 20~$\mu$m.  The interelectrode conduction is
dominated by the channel region due to this geometry.  No annealing
was performed following electrode deposition.


\section{Addendum}
  
\noindent{\bf Supplementary Information} is linked to the online version of the paper.\\

\noindent{\bf Acknowledgements}.  This work was supported by the US Department
  of Energy grant DE-FG02-06ER46337.  DN also acknowledges the David
  and Lucille Packard Foundation and the Research Corporation.  VLC
  acknowledges the NSF Center for Biological and Environmental
  Nanotechnology (EEC-0647452), Office of Naval Research
  (N00014-04-1-0003), and the US Environmental Protection Agency Star
  Program (RD-83253601-0).  CTY acknowledges a Robert A. Welch
  Foundation (C-1349) graduate fellowship.  RGSS and IVS acknowledge
  the Science Foundation Ireland grant 06/IN.1/I91.\\

\noindent{\bf Author Contributions}.   SL fabricated and measured the devices in this work and analyzed the data.  AF fabricated devices and performed XRD characterization of the nanocrystal materials.  DN and SL wrote the paper.  JTM and CYZ made the nanocrystals in VLC's laboratory, and VLC contributed expertise in nanomaterials chemistry and characterization.  RGSS and IVS grew the magnetite films, and IVS contributed expertise on magnetite physical properties.  All authors discussed the results and commented on the manuscript. \\

\noindent{\bf Competing Interests}.  The authors declare that they have no competing financial interests.\\

\noindent{\bf Correspondence}.  Correspondence and requests for materials should be addressed to D.N.~(email: natelson@rice.edu).

\end{document}